\input harvmac
%\draft
\noblackbox
%-------------------------
% This paper uses harvmac
%-------------------------
\font\ticp=cmcsc10

%
%-------------------
%  definitions
%-------------------

%
\def\half{{1\over2}}

\def\a{\alpha}

%
%-------------------
% references
%
\lref\spn{J. Breckenridge, R. Myers, A. Peet and C. Vafa, hep-th/9602065.}
\lref\ascv{A. Strominger and C. Vafa, hep-th/9601029.}
\lref\cama{C. Callan and J. Maldacena, hep-th/9602043.}
\lref\ghas{G. Horowitz and A. Strominger, hep-th/9602051.}
\lref\pol{J. Polchinski, hep-th/9510017.}
\lref\witb{E. Witten, hep-th/9510135.}
\lref\vgas{C. Vafa, hep-th/9511088.}
\lref\bsv{M. Bershadsky, V. Sadov and C. Vafa,
hep-th/9511222.}
\lref\vins{C. Vafa, hep-th/9512078.}
\lref\cvetd{M. Cvetic and D. Youm, hep-th/9507090.}
\lref\chrs{D. Christodolou, Phys. Rev. Lett. {\bf 25}, (1970) 1596;
D. Christodolou and R. Ruffini, Phys. Rev. {\bf D4}, (1971) 3552.}
\lref\cart{B. Carter, Nature {\bf 238} (1972) 71.}
\lref\penr{R. Penrose and R. Floyd, Nature {\bf 229} (1971) 77.}
\lref\hawka{S. Hawking, Phys. Rev. Lett. {\bf 26}, (1971) 1344.}
\lref\sussc{L.~Susskind,  Phys. Rev. Lett. {\bf 71}, (1993) 2367;
L.~Susskind and L.~Thorlacius, Phys. Rev. {\bf D49} (1994) 966;
L.~Susskind, ibid.  6606.}
\lref\polc{J. Dai, R. Leigh and J. Polchinski, Mod. Phys.
Lett. {\bf A4} (1989) 2073.}
\lref\hrva{P. Horava, Phys. Lett. {\bf B231} (1989) 251.}
\lref\cakl{C. Callan and I. Klebanov, hep-th/9511173.}
\lref\prskll{J. Preskill, P. Schwarz, A. Shapere, S. Trivedi and
F. Wilczek, Mod. Phys. Lett. {\bf A6} (1991) 2353. }
\lref\sbg{S. Giddings, Phys. Rev {\bf D49} (1994) 4078.}
\lref\cghs{C. Callan, S. Giddings, J. Harvey, and A. Strominger,
Phys. Rev. {\bf D45} (1992) R1005.}
\lref\bhole{G. Horowitz and A. Strominger,
Nucl. Phys. {\bf B360} (1991) 197.}
\lref\bekb{J. Bekenstein, Phys. Rev {\bf D12} (1975) 3077.}
\lref\hawkb{S. Hawking, Phys. Rev {\bf D13} (1976) 191.}
\lref\wilc{P. Kraus and F. Wilczek, hep-th/9411219, Nucl. Phys.
{\bf B433} (1995) 403. }
\lref\intrp{G. Gibbons and P. Townsend, Phys. Rev. Lett.
{\bf 71} (1993) 3754.}
\lref\gmrn{G. Gibbons, Nucl. Phys. {\bf B207} (1982) 337;
G. Gibbons and K. Maeda Nucl. Phys. {\bf B298} (1988) 741.}
\lref\bch{J. Bardeen, B. Carter and S. Hawking,
Comm. Math. Phys. {\bf 31} (1973) 161.}
\lref\mdr{W. Zurek and K. Thorne, Phys. Rev. Lett. {\bf 54}, (1985) 2171.}
\lref\stas{A.~Strominger and S.~Trivedi,  Phys.~Rev. {\bf D48}
 (1993) 5778.}
\lref\jpas{J.~Polchinski and A.~Strominger,
hep-th/9407008, Phys. Rev. {\bf D50} (1994) 7403.}
\lref\send{A. Sen, hep-th/9510229, hep-th/9511026}
\lref\cvet{M. Cvetic and A. Tseytlin, hep-th/9512031.}
\lref\kall{R. Kallosh, A. Linde, T. Ortin, A. Peet and
A. van Proeyen, Phys. Rev. {\bf D46} (1992) 5278.}
\lref\lawi{F. Larsen and F. Wilczek, hep-th/9511064.}
\lref\bek{J. Bekenstein, Lett. Nuov. Cimento {\bf 4} (1972) 737,
Phys. Rev. {\bf D7} (1973) 2333, Phys. Rev. {\bf D9} (1974) 3292.}
\lref\hawk{S. Hawking, Nature {\bf 248} (1974) 30, Comm. Math. Phys.
{\bf 43} 1975.}
\lref\sen{A. Sen, hep-th/9504147.}
\lref\suss{L. Susskind, hep-th/9309145.}
\lref\sug{L. Susskind and J. Uglum, hep-th/9401070,Phys. Rev. {\bf D50}
 (1994) 2700.}
\lref\peet{A. Peet, hep-th/9506200.}
\lref\tei{C. Teitelboim, hep-th/9510180.}
\lref\carl{S. Carlip, gr-qc/9509024. }
\lref\thoo{G. 'tHooft, Nucl. Phys. {\bf B335} (1990) 138;
Phys. Scr. {\bf T36} (1991) 247.}
\lref\fks{S. Ferrara, R. Kallosh and A. Strominger, hep-th/9508072,
Phys. Rev. {\bf D 52}, (1995) 5412 .}
%-------------------
% title page
%-------------------
%
\baselineskip 12pt
\Title{\vbox{\baselineskip12pt
\line{\hfil HUTP-96/A008, McGill/96-10,
PUPT-1599, UCSBTH-96-04}
\line{\hfil \tt hep-th/9603078} }}
{\vbox{\centerline{\bf MACROSCOPIC AND MICROSCOPIC ENTROPY OF}
\centerline{\bf NEAR-EXTREMAL SPINNING BLACK HOLES}}}
\vskip .1in
\centerline{\ticp J.C.  Breckenridge$^\dagger$, D.A. Lowe$^\ddagger$,
R.C. Myers$^\dagger$, }
\centerline{\ticp A.W. Peet$^{\natural}$, A. Strominger$^\ddagger$ and C.
Vafa$^{\S}$}
\bigskip
\centerline{$^\dagger$\it Physics Department, McGill University,
Montreal, PQ, H3A 2T8, Canada}
\vskip.1in
\centerline{$^\ddagger$\it Department of Physics, University of California,
Santa Barbara, CA 93106, USA}
\vskip.1in
\centerline{$^{\natural}$\it Joseph Henry Laboratories, Princeton University,
Princeton, NJ 08544, USA}
\vskip.1in
\centerline{$^{\S}$\it Lyman Laboratory of Physics, Harvard University,
Cambridge, MA 02138, USA}
\bigskip
\centerline{\bf Abstract}
A seven parameter
family of five-dimensional black hole solutions
depending on mass, two angular momenta,
three charges and the asymptotic
value of a scalar field is constructed. The entropy is computed as
a function of these parameters both from the
Bekenstein-Hawking formula and from the degeneracies of the corresponding
D-brane states in string theory. The expressions agree at and to
leading order away from extremality.
\Date{March 1996}
%
%----------------------
% Body of Paper

\newsec{Introduction}

In recent months string theory has demonstrated a remarkable and
detailed knowledge of black hole thermodynamics. In \ascv\ it was found
that the newly-understood rules \refs{\polc \hrva \witb \send \vgas \bsv
{--}\vins}
for counting degeneracies of BPS-saturated, D-brane
soliton bound states precisely reproduces the Bekenstein-Hawking entropy
for a certain five-dimensional extremal Reissner-Nordstrom black hole.
These results were extended to leading order above extremality in
\refs{\cama,\ghas}, solidifying the identification of
the microscopic states responsible for
the entropy.  In \spn\
it was shown that the stringy
degeneracies continue to match the extremal
Bekenstein-Hawking entropy when rotation
is added.
Ordinarily the addition of rotation (without energy)
to an extremal Reissner-Nordstrom black hole
destabilizes the horizon and yields a naked singularity. However
string theory,
in order to avoid a conflict with the microscopic counting,
cleverly stabilizes the horizon with the help of
a Chern-Simon coupling in the low-energy field theory.
In the process a qualitatively new class of
supersymmetric spinning black hole solutions was found\spn.

In this paper we will combine the analyses of \refs{\ghas, \spn}
and consider these new spinning solutions just above extremality. Again
we will find
perfect agreement - a seven parameter fit -
between the detailed thermodynamic behavior predicted by
the Bekenstein-Hawking entropy and by the microscopic state counting.

\newsec{The Rotating Nonextremal Black Hole}

The low-energy action for six-dimensional type IIB string theory
contains the terms
\eqn\fds{{1\over 16 \pi}
\int d^6x \sqrt{- g} ( R-(\nabla \phi )^2
-{1 \over 12} e^{2\phi}H^2 )}
in the six-dimensional Einstein frame. $H$ denotes the RR three form field
strength.
We adopt conventions in
which $G_N=1$. The scalar $\phi$ here is the logarithm of the
volume of the internal four-manifold in the string frame.
The ten-dimensional string dilaton is an arbitrary constant for our solutions
and is suppressed.
We will further compactify to five dimensions by
periodically identifying $y \sim y+L$, where
$y$ denotes the fifth spatial coordinate.
We take the asymptotic
length $L$ of the compact dimension to be very large.

The solutions of
interest to us are most simply represented as six-dimensional
black string
solutions to \fds, which wind around the $y$ direction and
hence are black holes in five dimensions.
The six-dimensional black string can can carry both electric
and magnetic charge with respect to
$H$ :
\eqn\qhd{\eqalign{Q_+&\equiv {1\over 8}\int_{S^3} e^{2\phi}\ *H ,\cr
 Q_- &\equiv {1\over 4\pi^2}\int_{S^3} \ H ~ .\cr}}
It may also carry total ADM momentum $P$ along the $y$ direction which appears
in
five dimensions
as an electric charge \ghas:
\eqn\mmt{P\equiv {2  \pi n \over L}~.}
We have chosen our conventions so that $n$ and $Q_-Q_+ \equiv \half Q^2$ are
integers.
In five spacetime dimensions the spatial rotation group is $SU(2) \times
SU(2)$. Hence solutions are in addition labeled by
two angular momenta.

Black string solutions are also characterized by the asymptotic value of
$\phi$. We are primarily interested in the
entropy which cannot depend on the asymptotic value of
$\phi$ \refs{\fks \cvetd \cvet {--}\lawi}. For a special asymptotic value
$\phi_h$, the sources for $\phi$
cancel exactly and the equations of motion imply $\phi$ is
constant everywhere. This special value is
\eqn\qhds{
e^{2\phi_h}= {{2 Q_+}\over { \pi^2 Q_-}} ~.
}
In order to compute the entropy it is sufficient to consider the solutions
with $\phi=\phi_h$.

Reduction from six to five dimensions yields a second
five dimensional scalar field whose asymptotic value is
$L$, the size of the $S^1$ parameterized by $y$.
This scalar could also be frozen to a value which would be proportional to
$n/Q$. However it is important not to freeze this field because we
will need to compute how the entropy varies as a function of both the
energy and $n$ with all other quantities - in particular the asymptotic
values of the fields - held fixed. This is impossible to do if the
value of the scalar field is tied to $n/Q$.  This problem does not arise
for the scalar $\phi$ because, once the behavior of the entropy is
known for any value of the ratio $Q_+/Q_-$, it is determined for any other
value by duality which implies that it can depend only on the product
$Q^2/2$.

The solutions of interest can be generated by
beginning with the
five dimensional Kerr solution which spins in two
independent planes.
This may be lifted to
a black string solution of heterotic string theory
in six dimensions by adding
a trivial flat direction $y$. A boost is performed
mixing the time
direction $t$ with an internal direction to yield
a nontrivial right-handed gauge field. Next, by
applying
string-string duality followed by a T-duality transformation, one obtains a
black string
solution of Type IIB
string theory in six dimensions. Lastly, a boost is
performed along the string yielding the following
solution
\vfill
\eject
\eqn\met{\eqalign{
ds_6^2 &=
- \left[1-{{(r_+^2\cosh^2\alpha-r_-^2\sinh^2\alpha)}
  \over{\rho^2}}\right] dt^2
+ \left[1-{{(r_-^2\cosh^2\alpha-r_+^2\sinh^2\alpha)}
  \over{\rho^2}}\right] dy^2 \cr
&\qquad
+ \sin^2\theta \left[r^2+a^2+{{(a^2r_+^2-b^2
r_-^2)\sin^2\theta}\over{\rho^2}}\right]
  d\varphi^2  \cr & \qquad
+ \cos^2\theta
  \left[r^2 +b^2+ {{(b^2 r_+^2-a^2r_-^2)\cos^2\theta}\over{\rho^2}}\right]
d\psi^2 \cr &\qquad
 + \rho^2 d\theta^2 +
{{\rho^2}\over{r^2}}\biggl[ \left(1-{{r_-^2}\over{r^2}}\right)
  \left(1-{{r_+^2-a^2-b^2}\over{r^2}}\right) + {a^2 b^2 \over r^4} \biggr]^{-1}
dr^2 \cr
&\qquad
+ 2\sin^2\theta{{(ar_+^2 \cosh\alpha- b r_-^2 \sinh \alpha) }\over{\rho^2}}dt
d\varphi \cr & \qquad
+ 2\sin^2\theta{{(ar_+^2\sinh\alpha- b r_-^2 \cosh \alpha) }\over{\rho^2}}dy
d\varphi \cr
&\qquad + 2\cos^2\theta{{( b r_+^2 \cosh \alpha - ar_-^2\sinh\alpha)
}\over{\rho^2}}dt d\psi \cr
& \qquad
+ 2\cos^2\theta{{(br_+^2 \sinh \alpha - ar_-^2\cosh\alpha)
}\over{\rho^2}}dy d\psi \cr
&\qquad
+ 2\cosh\alpha\sinh\alpha{{(r_+^2-r_-^2)}\over{\rho^2}}dt dy + 2 \cos^2\theta
\sin^2 \theta  {{a b (r_+^2-r_-^2)}\over \rho^2} d \varphi d\psi~,\cr
\phi&=\phi_h~,\cr
}}
where
$\rho^2 \equiv r^2+a^2\cos^2 \theta+ b^2
 \sin^2\theta$,
$\a$ is a boost parameter, and $a$,$b$ are components of the angular
momentum per unit mass of the original Kerr
solution. A nontrivial RR three-form field strength
is present, but its precise form will not be needed
in the following.
The parameters $r_\pm$  are related to the charge
by $Q^2\equiv 2 Q_+ Q_- = (\pi r_+  r_-)^2 $.
The outer and inner event horizons are located at
\eqn\ehor{
r^2 = {1\over 2}\biggl( r_+^2 +r_-^2-a^2-b^2\pm \sqrt{
(r_+^2-r_-^2-a^2-b^2)^2-4a^2 b^2} \biggr)~.
}

The six-dimensional ADM energy of this solution is
\eqn\admmass{
E ={ {L \pi}\over 8} (2(r_+^2 + r_-^2) + (r_+^2 - r_-^2) \cosh 2 \alpha )~.
}
The ADM momentum along the string is given by
\eqn\momentum{
P =  {{\pi L}\over 8} \sinh 2\alpha
(r_+^2 -r_-^2)~.
}
The angular momenta in the independent planes
defined by $\varphi$ and $\psi$ are
\eqn\angmom{
\eqalign{
J_1\equiv J_\varphi &= {{\pi  L }\over 4} (a r_+^2 \cosh \alpha - b r_-^2 \sinh
\alpha)
{}~, \cr
J_2\equiv J_\psi  &=  {{\pi  L }\over 4}(b r_+^2 \cosh \alpha - a r_-^2 \sinh
\alpha)~.\cr}
}

Following \ghas, we expect the Bekenstein-Hawking entropy to agree with the
D-brane counting away
from the supersymmetric extremal limit \spn\  provided the momentum density
$P/L$
and the excitation energy density $\delta E/L$ are
small.
To study this limit we expand
\eqn\radpm{
r_\pm = r_0 \pm \epsilon~,
}
with $\epsilon \ll 1$, and $\alpha$ finite.
Note we need to take the limit in such a way
that $r_+^2 -(|a|+|b|)^2 > r_-^2$
in order
to avoid naked singularities. This implies that
$a^2$ and $b^2$ are of order $\epsilon$. The
longitudinal size of the string near the horizon
is finite in this limit.
To first order in $\epsilon$, the excitation energy
is
\eqn\exciten{
\delta E = {{L \pi r_0 \epsilon}\over 2} \cosh 2\alpha~,
}
and the entropy is
\eqn\nonentropy{
\eqalign{
S &= {1\over 2} L \pi^2 r_0^2 \biggl( (4 r_0 \epsilon-
a^2-b^2)\cosh^2 \alpha + 2 a b \cosh\alpha \sinh \alpha - {1\over 2}(4 r_0
\epsilon -a^2-b^2) \cr &
 +
{1\over 2} \sqrt{(4 r_0 \epsilon -a^2-b^2 -2 ab)(4
r_0 \epsilon -a^2-b^2 +2ab)} \biggr)^{1/2}~. \cr}
}
Now define the following quantities
\eqn\oscnum{
\eqalign{
\tilde n_R &= {L \over {4\pi}} (\delta E + P)-
{ {(J_1-J_2)^2}\over {2 Q^2} } ~,\cr
\tilde n_L &= {L \over {4\pi}} (\delta E - P)-
{ {(J_1+J_2)^2}\over {2 Q^2} } ~.\cr}
}
The entropy \nonentropy\ may then be expressed as
\eqn\niceent{
S= \pi Q( \sqrt{2 \tilde n_L} + \sqrt{2 \tilde n_R})
{}~.
}

\newsec{Matching to D-brane Degeneracies}

As discussed in \ascv\ the $P=0$ black hole ground state is
a bound state of $Q_+$ RR onebranes wound around the
$S^1$ in the $y$ direction with $Q_-$ RR fivebranes wound
around both the $S^1$ and the internal four-manifold.  In the limit
of large radius $L$ for $S^1$, the excitations of this system
are described by a supersymmetric
sigma model on a manifold of real dimension
$2Q^2$ \refs{\vgas \bsv
{--}\vins} .   In the regime of charges we are interested in,
to leading order the degeneracy comes from string modes
with short wavelengths and hence the curvature
of the manifold is irrelevant.  Thus we have
the same leading degeneracy as the
excitations of $2Q^2$ species of massless
bosons and $2Q^2$ species
of massless fermions which move around the $S^1$.
 Ignoring angular momentum, the entropy of
$N_B$ ($N_F$) species of right-moving bosons (fermions)
with total energy $E_R$ in a box of length $L$ is given by the
standard thermodynamic formula
\eqn\est{S=\sqrt{\pi (2N_B+N_F)E_RL\over 6}~.}
At low energies and large $L$ the system is dilute, interactions can
be ignored, and the entropy is additive. Hence, using $N_F=N_B=2Q^2$
and $E_{R,L}=2\pi n_{R,L}/L$,
\est\ becomes
\refs{\ascv,\ghas}
\eqn\ssr{S= \pi Q( \sqrt{2 n_L} + \sqrt{2 n_R})~,}
where $n_{L,R}$ are given by \oscnum\ with $J_1=J_2=0$.

Now let us correct for the angular momentum. As argued in
\spn\ $J_1+J_2$ is carried by left-movers, while $J_1-J_2$
is carried by right movers. Fixing the total angular momentum
carried by the right movers decreases the number of states
available for fixed energy. As shown in \spn\ the effect of
this on the entropy for left-movers only is to replace
$n_L$ with $\tilde n_L$. However since the entropy of
left and right movers is additive we have simply
\eqn\nicent{
S= \pi Q( \sqrt{2 \tilde n_L} +
\sqrt{2 \tilde n_R})~,
}
in agreement with the black hole calculation \niceent.

\bigskip
\centerline{\bf Acknowledgements}

We would like to thank R. Emparan and
G. Horowitz for discussions.
The research of J.C.B and R.C.M is supported by
NSERC of Canada and Fonds FCAR du Qu\'ebec; that
of D.A.L by NSF grant PHY-91-16964;
that of A.W.P. by NSF grant PHY-90-21984; that of
A.S. by DOE grant DOE-91ER40618;
and that of C.V. by NSF grant PHY-92-18167.

\listrefs
\end